# Pseudo-time-reversal symmetry and topological edge states in two-dimensional acoustic crystals

*Jun Mei[1], Zeguo Chen[2], and Ying Wu[2,a)]*

[1]*Department of Physics, South China University of Technology, Guangzhou 510640, People's Republic of China*

[2]*Division of Computer, Electrical and Mathematical Sciences and Engineering, King Abdullah University of Science and Technology (KAUST), Thuwal 23955-6900, Saudi Arabia*

**Abstract**

We propose a simple two-dimensional acoustic crystal to realize topologically protected edge states for acoustic waves. The acoustic crystal is composed of a triangular array of core-shell cylinders embedded in a water host. By utilizing the point group symmetry of two doubly degenerate eigenstates at the $\Gamma$ point, we can construct pseudo-time-reversal symmetry as well as pseudo-spin states in this classical system. We develop an effective Hamiltonian model for the associated dispersion bands around the Brillouin zone center, and find the inherent link between the band inversion and the topological phase transition. With numerical simulations, we unambiguously demonstrate the unidirectional propagation of acoustic edge states along the interface between a topologically nontrivial acoustic crystal and a trivial one, and the robustness of the edge states against defects with sharp bends. Our work provides a new design paradigm for manipulating and transporting acoustic waves in a topologically protected manner. Technological applications and devices based on our design are expected in various frequency ranges of interest, spanning from infrasound to ultrasound.

[a)] Author to whom correspondence should be addressed. Electronic mail: ying.wu@kaust.edu.sa

Topological insulators have witnessed a lot of success in the last decade. In various electronic systems, from the artificially designed Haldane lattice [1], to graphene [2, 3] and HgTe/CdTe quantum well structures [4, 5], the topology – a mathematical property describing the quantized behavior of the wavefunctions over the associated dispersion bands – has been found to have a profound influence on the transportation properties of electronic wave functions [6, 7]. The concept of topology was thereafter borrowed from quantum systems and transplanted into classical systems, offering researchers a new degree of freedom in controlling and manipulating electromagnetic [8-23], acoustic [24-29] and elastic waves [30-38] in their corresponding artificial structures.

In two- or three-dimensional acoustic systems, the realization of topological nontrivial phases was reported either by rotating fluids to break the time-reversal symmetry [25-27], or by utilizing chiral interlayer coupling to break the inversion symmetry [28]. Although topologically protected *chiral* edge states were numerically demonstrated in these works, the need for integration of rotational fluids into resonators or the fabrication of complex inversion-breaking chiral structures remains technically. Different from chiral edge states, *helical* edge states belong to another kind of topological nontrivial phase with a different mechanism. In the quantum spin Hall effect (QSHE), the helical edge state is protected by the locking of its spin with its momentum and the existence of time reversal symmetry. Although helical edge states were studied in artificial electromagnetic [11, 15, 17, 19, 22, 23] and elastic structures [32, 35], relatively few works have described acoustic waves [29] with rather complex structures. Because acoustic waves are scalar waves and do not have polarization, realizing unidirectional propagation in acoustics is not trivial. Here, we propose a different and *simple* way to realize *helical* edge states in acoustic waves. The system is a two-dimensional acoustic crystal (AC) composed of a triangular array of core-shell cylinders embedded in a water host. We show that by utilizing the rotational symmetry of the unit cell, pseudo-time-reversal symmetry [23] can be constructed and, as a result, the acoustic analogue of the QSHE can be emulated. We unambiguously demonstrate topologically protected acoustic one-way edge states with robust propagation against scattering from defects.

A schematic of the two-dimensional (2D) AC is shown in Fig. 1(a). Each core-shell cylinder has a steel rod with radius $r$ as its core, which is coated by a layer of silicone rubber. The outer radius of the cylinder is $R$, and $a$ is the lattice constant. The acoustic wave equation is

$$\nabla \cdot \left( \frac{1}{\rho_r(\vec{r})} \nabla p \right) = -\frac{\omega^2}{c^2} \cdot \frac{p}{B_r(\vec{r})} \tag{1}$$

where $p$ is the pressure, with $\rho_r = \rho/\rho_0$ and $B_r = B/B_0$ being the relative mass density and bulk modulus, respectively. $c = \sqrt{B_0/\rho_0}$ is the speed of sound in water. The mass densities for water, rubber and steel are $\rho_0=1000 kg/m^3$, $\rho_1=1300 kg/m^3$, and $\rho_2=7670 kg/m^3$, respectively. The longitudinal wave velocities in water, rubber and steel are $c_0=1490 m/s$, $c_1=489.9 m/s$, and $c_2=6010 m/s$, respectively. Due to the strong mismatch between the longitudinal velocities in these media, the shear wave modes in the solid components are ignored here, and this simplification does not alter the essential physics of the system [39, 40].

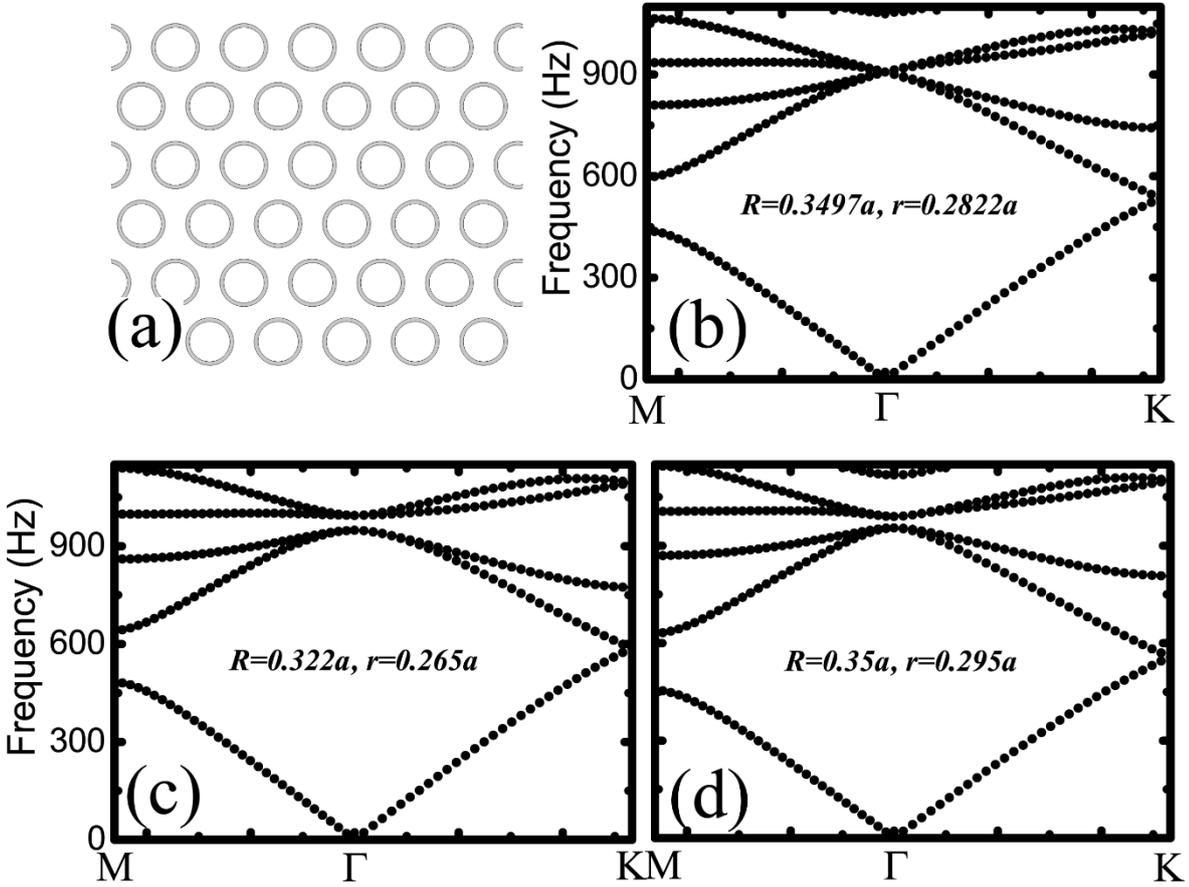

Figure 1. (a) The 2D AC consists of a triangular array of core-shell cylinders embedded in a water host. (b) The band structure for $r=0.2822a$ and $R=0.3497a$; a double Dirac cone is seen at the $\Gamma$ point. (c) The band structure for $r=0.265a$ and $R=0.322a$; a trivial gap is formed between the third and fourth bands. (d) The band structure for $r=0.295a$ and $R=0.35a$; a nontrivial gap is formed. The frequency range of the gap is almost the same in (c) and (d).

It was shown in Ref. [40] that a four-fold degeneracy is achieved at the Γ point at frequency $\omega_D=0.6092(2\pi c_0/a)$ when $r=0.2822a$ and $R=0.3497a$, and as a result a double Dirac cone was formed at the center of the Brillouin zone as shown in Fig. 1(b). Here, we use COMSOL Multiphysics, a commercial package based on the finite-element method, to calculate the band structures.

The four-fold degeneracy is realized when the doubly degenerate dipolar states coincide with the doubly degenerate quadrupolar states at $\omega_D$. The degeneracy is thus accidental. That is to say, if we alter the geometric parameters, e.g., the inner and/or outer radii of the core-shell cylinders, the four-fold degeneracy will be lifted, and the dipolar states will be separated from the quadrupolar states. In Figs. 1(c) and 1(d), we plot the band structures for $r=0.265a$ and $R=0.322a$, and for $r=0.295a$ and $R=0.35a$, respectively, where the four-fold degeneracy is lifted and the dipolar states are separated from the quadrupolar states.

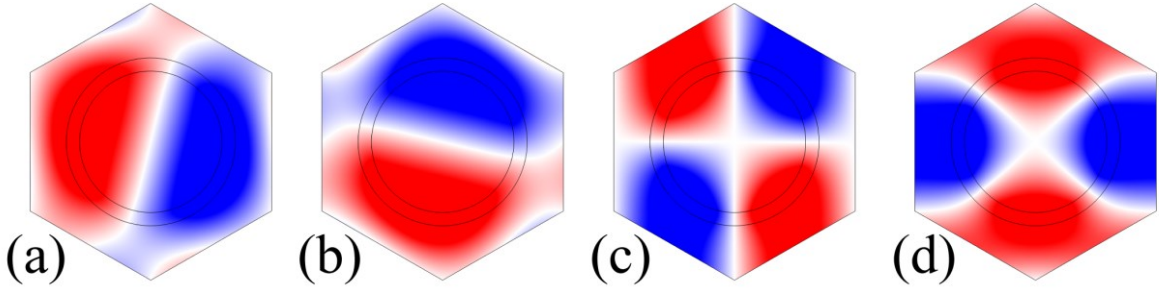

Figure 2. Pressure field distributions of the $E_1$ and $E_2$ representations for the AC shown in Fig. 1(d). In (a) and (b), $p_x$- and $p_y$-like pressure fields are seen, while in (c) and (d) $d_{x^2-y^2}$- and $d_{2xy}$-like patterns are recognized. Red and blue denote the positive and negative maxima, respectively.

For the AC we mentioned above, the point group at the center of the Brillouin zone (BZ) is $C_{6v}$, which has two 2D irreducible representations: $E_1$ with basis functions $(x,y)$ and $E_2$ with basis functions $(2xy, x^2-y^2)$ [41]. $E_1$ modes have odd spatial parity, while $E_2$ modes have even spatial parity. From the distribution of the eigenfields shown in Fig. 2, it is easy to recognize that the $E_1$ [Figs. 2(a) & 2(b)] and $E_2$ [Figs. 2(c) & 2(d)] representations have, respectively, the same symmetry as the $(p_x, p_y)$ and $(d_{x^2-y^2}, d_{2xy})$ orbitals of electrons in quantum systems. We note that relative eigenfrequencies corresponding to different irreducible representations change as the geometry of the core-shell cylinder changes. To be more specific, the eigenfrequency associated with the $E_1$ representation is lower than that of $E_2$ as shown in Fig. 1(c), but it is higher in Fig. 1(d), indicating that a band inversion process occurs as the geometry changes. Later, we will show that this band inversion is inherently associated with a topological phase transition.

In the following, we demonstrate that the spatial symmetry of the $E_1$ and $E_2$ representations can be utilized to construct pseudo-time-reversal symmetry [23]. Let $D^{E_1}(C_6)$ and $D^{E_1}(C_6^2)$ denote the $E_1$ irreducible representations of the $\pi/3$ and $2\pi/3$ rotations, respectively. Their matrix representations on basis $(x, y)^T$ are therefore

$$D^{E_1}(C_6) = \begin{pmatrix} 1/2 & -\sqrt{3}/2 \\ \sqrt{3}/2 & 1/2 \end{pmatrix} \tag{2}$$

$$D^{E_1}(C_6^2) = \begin{pmatrix} -1/2 & -\sqrt{3}/2 \\ \sqrt{3}/2 & -1/2 \end{pmatrix}. \tag{3}$$

It turns out that they can be combined together in a unitary operator, $U$, as

$$U = \left[ D^{E_1}(C_6) + D^{E_1}(C_6^2) \right]/\sqrt{3} = \begin{pmatrix} 0 & -1 \\ 1 & 0 \end{pmatrix} = -i\sigma_y, \tag{4}$$

where $\sigma_y$ is the Pauli matrix. Obviously, $U^2 = -I$. Thus, we can construct an anti-unitary operator, $T$, as $T = UK = -i\sigma_y K$, where $K$ is the complex conjugate operator. It follows that

$$T^2 \begin{pmatrix} p_x \\ p_y \end{pmatrix} = T \begin{pmatrix} -p_y \\ p_x \end{pmatrix} = -\begin{pmatrix} p_x \\ p_y \end{pmatrix}, \tag{5}$$

which yields $T^2 = -I$. Similarly, the $E_2$ matrix representations of rotational operators $C_6$ and $C_6^2$ on basis $(x^2 - y^2, 2xy)^T$ are

$$D^{E_2}(C_6) = \begin{pmatrix} -1/2 & -\sqrt{3}/2 \\ \sqrt{3}/2 & -1/2 \end{pmatrix} \tag{6}$$

$$D^{E_2}(C_6^2) = \begin{pmatrix} -1/2 & \sqrt{3}/2 \\ -\sqrt{3}/2 & -1/2 \end{pmatrix}. \tag{7}$$

And the unitary operator, $U$, can be constructed as $U = \left[ D^{E_2}(C_6) - D^{E_2}(C_6^2) \right]/\sqrt{3} = -i\sigma_y$. It is easy to check that we also have $T^2 = -I$, the same as in the $E_1$ irreducible representation. Therefore, for both $E_1$ and $E_2$ modes, the pseudo-time-reversal symmetry, $T = UK$, in the current acoustic system indeed satisfies $T^2 = -I$, which is similar to the *real* time-reversal symmetry in electronic systems and which also guarantees the appearance of a Kramers doublet at the $\Gamma$ point. From the derivations shown above, it is clear that the role played by the crystal symmetry of the unit cell in constructing the pseudo-time-reversal symmetry is crucial.

Following this analysis, we can construct the pseudo-spin states as $p_\pm = (p_x \pm ip_y)/\sqrt{2}$, with $p_+$ ($p_-$) being the pseudo spin-up (spin-down) state. On the $(p_+, p_-)^T$ basis, the pseudo-time-reversal operator, $T' = U'K$, [$U'$ and $T'$ are defined on the $(p_+, p_-)^T$ basis, while $U$ and $T$ are defined on the $(p_x, p_y)^T$ basis] exhibits the following desired properties:

$$\begin{cases} T'p_+ = -ip_- \\ T'p_- = ip_+ \end{cases} \quad \begin{cases} T'^2 p_+ = -p_+ \\ T'^2 p_- = -p_- \end{cases} \tag{8}$$

From Eq. (8), it is clear that the wave functions $(p_+, p_-)$ are the two pseudo-spin states in the $E_1$ representation of our acoustic system because the pseudo-time-reversal operator, $T'$, transforms the pseudo spin-up state into a spin-down state, and vice versa. The same conclusion can be made on $d_\pm = (d_{x^2-y^2} \pm id_{2xy})/\sqrt{2}$, which is another pair of pseudo spin-up/spin-down states associated with the $E_2$ representation.

To understand the topological property of the band gaps shown in Figs. 1(b) and 1(c), we construct an effective Hamiltonian for the current system around the $\Gamma$ point from a $\vec{k} \cdot \vec{p}$ perturbation method [40-42]. We assume $\Gamma_\alpha (\alpha = 1,2,3,4)$ are the four eigenstates at the $\Gamma$ point: $\Gamma_1 = p_x$, $\Gamma_2 = p_y$, $\Gamma_3 = d_{x^2-y^2}$, and $\Gamma_4 = d_{2xy}$, with the $p_{x(y)}$ and $d_{x^2-y^2(2xy)}$ states corresponding to the $E_1$ and $E_2$ representations, respectively. According to the degenerate second-order perturbation theory [41], the effective Hamiltonian around the $\Gamma$ point is given by $H_{mn}^{eff} = H'_{mn} + \sum_\alpha \frac{H'_{m\alpha} H'_{\alpha n}}{\varepsilon_m^{(0)} - \varepsilon_\alpha^{(0)}}$ $(m,n = 1,2,3,4)$, where $\varepsilon_{1,2}^{(0)} = \varepsilon_p^0$ ($\varepsilon_{3,4}^{(0)} = \varepsilon_d^0$) is the eigenfrequency of $\Gamma_{1,2}$ ($\Gamma_{3,4}$), and $H' = \frac{2i}{\rho_r}\vec{k}\cdot\nabla + i\vec{k}\cdot\nabla\frac{2}{\rho_r} - \frac{2}{\rho_r}$ is the $\vec{k}\cdot\vec{p}$ perturbation term for the acoustic wave equation (1), obtained by expanding the Bloch eigenstates at point $\vec{k}$ as the linear combinations of the Bloch eigenstates at point $\Gamma$ [40]. Rewriting the above Hamiltonian on the basis $[p_+, d_+, p_-, d_-]$, we arrive at the following effective Hamiltonian in the vicinity of the $\Gamma$ point,

$$H^{eff}(\vec{k}) = \begin{pmatrix} M - Bk^2 & Ak_+ & 0 & 0 \\ A^*k_- & -M + Bk^2 & 0 & 0 \\ 0 & 0 & M - Bk^2 & Ak_- \\ 0 & 0 & A^*k_+ & -M + Bk^2 \end{pmatrix}, \quad (9)$$

where $k_\pm = k_x \pm ik_y$ and $M = \frac{\varepsilon_d^0 - \varepsilon_p^0}{2}$ is the frequency difference between $E_2$ and $E_1$ representations at the $\Gamma$ point, which is positive (negative) before (after) the band inversion. $B$ is determined by the diagonal elements of the second-order perturbation term $H'_{m\alpha}H'_{\alpha n}$, and is typically negative. $A$ comes from off-diagonal elements of the first-order perturbation term $H'_{mn} = \langle \Gamma_m | H' | \Gamma_n \rangle$ with $m = 1,2$ and $n = 3,4$. We note that to derive the above Hamiltonian, the spatial symmetries of the eigenstates, $\Gamma_\alpha$, are utilized. The effective Hamiltonian, $H^{eff}(\vec{k})$, shown in Eq. (9) takes a similar form as that proposed in the Bernevig-Hughes-Zhang (BHZ) model for the CdTe/HgTe/CdTe quantum well system [4].

We note that, in Fig. 1(c), the bands above (below) the gap belong to the $E_2$ ($E_1$) representation, which means that $M = \frac{\varepsilon_d^0 - \varepsilon_p^0}{2} > 0$. For the Hamiltonian expressed in Eq. (9), the spin Chern numbers can be evaluated as [4, 43]

$$C_\pm = \pm\frac{1}{2}[\text{sgn}(M) + \text{sgn}(B)] \quad (10)$$

Since $C_\pm = 0$, we conclude that the band gap shown in Fig. 1(c) is trivial. However the situation is different in Fig. 1(d), where the bands above (below) the gap exhibit $E_1$ ($E_2$) characteristics around the $\Gamma$ point, meaning that $M < 0$. Applying Eq. (10), we immediately know that $C_\pm = \pm 1$ and the gap in Fig. 1(d) is nontrivial. It is interesting to find the topological phase transition from a trivial one [Fig. 1(c)] to a nontrivial one [Fig. 1(d)] that is associated with the band inversion between the $E_1$ and $E_2$ representations around the $\Gamma$ point. Our finding shares a similar physical mechanism with that in the BHZ model developed for the CdTe/HgTe/CdTe quantum well system [4]. Therefore, we expect that our AC system can support an acoustic 'spin Hall effect', although our system is quite different from the BHZ quantum system: we are dealing with a system governed by the classical acoustic wave equation rather than the Schrödinger equation.

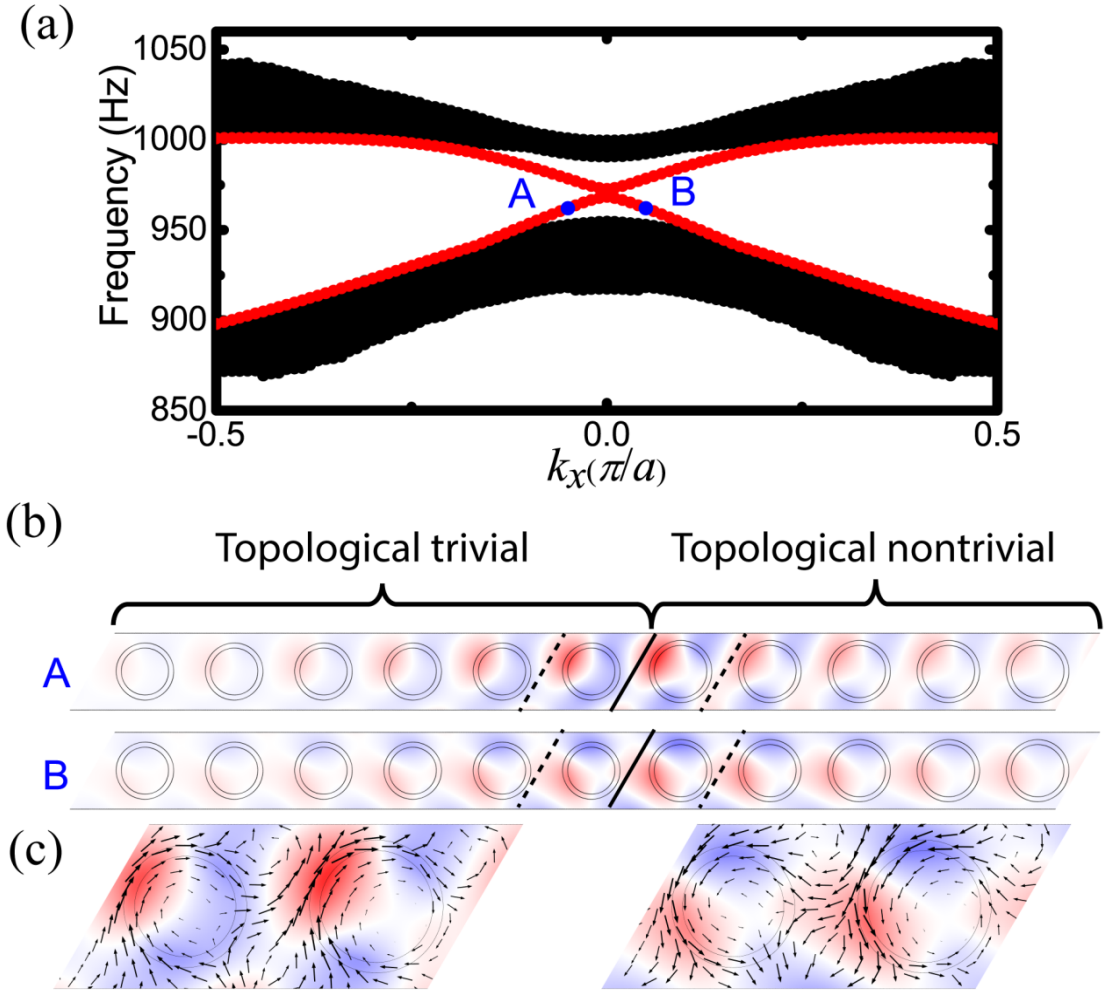

Figure 3. (a) The projected band structure along the $\Gamma K$ direction for a ribbon of topologically nontrivial crystal with its two edges cladded by topologically trivial crystals. The ribbon has 1 unit cell in one direction and 86 unit cells along the other direction (46 nontrivial unit cells cladded by 20 trivial unit cells on both sides). The red and black dots represent edge and bulk states, respectively. (b) & (c) Pressure field distributions around the interface between the trivial and nontrivial phases at points A and B, i.e., at $k_x = -0.05\pi/a$ and $0.05\pi/a$, respectively. Red and blue denote positive and negative maxima, respectively. Black arrows indicate the time-averaged Poynting vector.

We consider a ribbon of topologically nontrivial crystal (i.e., the AC that produces the band structure shown in Fig. 1(d)) with its two edges cladded by two topologically trivial crystals (the AC that produces the band structure shown in Fig. 1(c)). The frequency regime, [955.18 Hz, 990.43 Hz], is common for the trivial and nontrivial gaps to create true edge states that are spatially confined around the interface between two crystals. In Fig. 3(a), we plot the projected band structures along the $\Gamma K$ direction for such a ribbon. We find that in addition to the bulk states represented by black dots, there are doubly degenerate states, represented by red dots within the bulk gap region. After examining the eigenfield distributions at the red dots (e.g., points A and B), we find that the pressure field decays exponentially into bulk crystals on both sides, which means that the red curves represent the dispersion relations of edge states that are tightly confined around the interface between the nontrivial and trivial phases. In Fig. 3(b), we plot the pressure field distribution on one interface for the eigenstates at points A and B, respectively, and a magnified view is plotted in Fig. 3(c), where black arrows indicate the time-averaged Poynting vectors. The clockwise and anticlockwise distributions of the Poynting vector at the interface unveil the characteristics of the pseudo spin-up and spin-down states, respectively. The locking of the (pseudo) spin-up and spin-down states with counter-propagations of edge states is reminiscent of the quantum spin Hall effect in electronic systems.

Because the pseudo-time-reversal symmetry and the pseudo-spin states are constructed on the basis of the $C_{6v}$ point group symmetry, any deviation from the crystal symmetry would mix the two pseudo-spin channels as in other topological systems [11, 15, 23]. Actually, there is a tiny gap (not evident in Fig. 3(a)) at the $\Gamma$ point, arising from the reduction of the $C_{6v}$ symmetry at the interface between the trivial and nontrivial phases. However, the topological properties of the corresponding structures remain valid even with a moderate deformation in the lattice symmetry, as will be explicitly shown below.

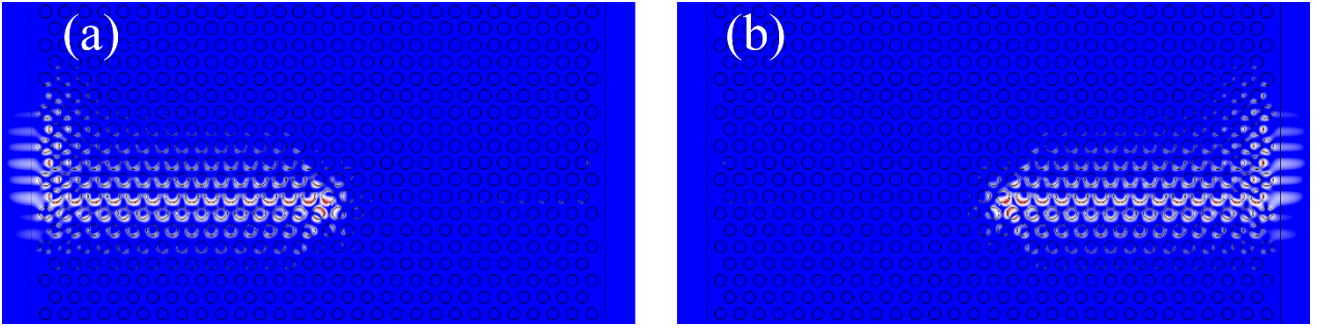

Figure 4. Realization of unidirectional acoustic wave propagation towards the (a) left and (b) right directions along the interface between a nontrivial phase (upper part) and a trivial phase (lower part), with the whole structure surrounded by perfectly matched layers. The acoustic waves that carry the leftward (rightward) wave vector is excited by using a couple of point sources located in the middle part of the interface at a distance of $\lambda_0/4$, and the two sources have a phase difference of $\pi/2$ ($-\pi/2$).

As a first example, we consider a flat edge between a topologically nontrivial phase (upper part) and a trivial phase (lower part), as shown in Fig. 4, where the whole structure is surrounded by perfectly matched layers (PMLs) to absorb outgoing waves. When an acoustic wave carrying a leftward (rightward) wave vector is excited in the middle part of the edge, unidirectional propagation of the acoustic wave towards the left (right) direction can be observed in Figs. 4(a) (4(b)). When these edge waves arrive at the left (right) boundary of the simulation domain, they are guided without reflections along the interface of the nontrivial crystal to continue propagating upwards. At the same time, they gradually decay into the PMLs. Negligible reflection occurs at the left (right) boundary of the edge, which is expected from the topological properties of the edge states.

One of the most important features of topological edge states is that they are immune to defects/imperfections. In the following, we demonstrate edge wave propagation around a specific type of imperfection: four sharp bends of the edge shown in Fig. 5 (b), which is constructed from the structure in Fig. 4 by further replacing a region of the nontrivial topological phase ($8\vec{u}_1 \times \circ \vec{u}_2$) with that of the trivial one. When a left-heading wave is excited, it propagates along the edge and can go around the rhombic defect without reflections at the four sharp corners. It also maintains its unidirectional propagation as shown in Fig. 5(a). This result confirms the topological robustness of the edge states against a sharply curved interface.

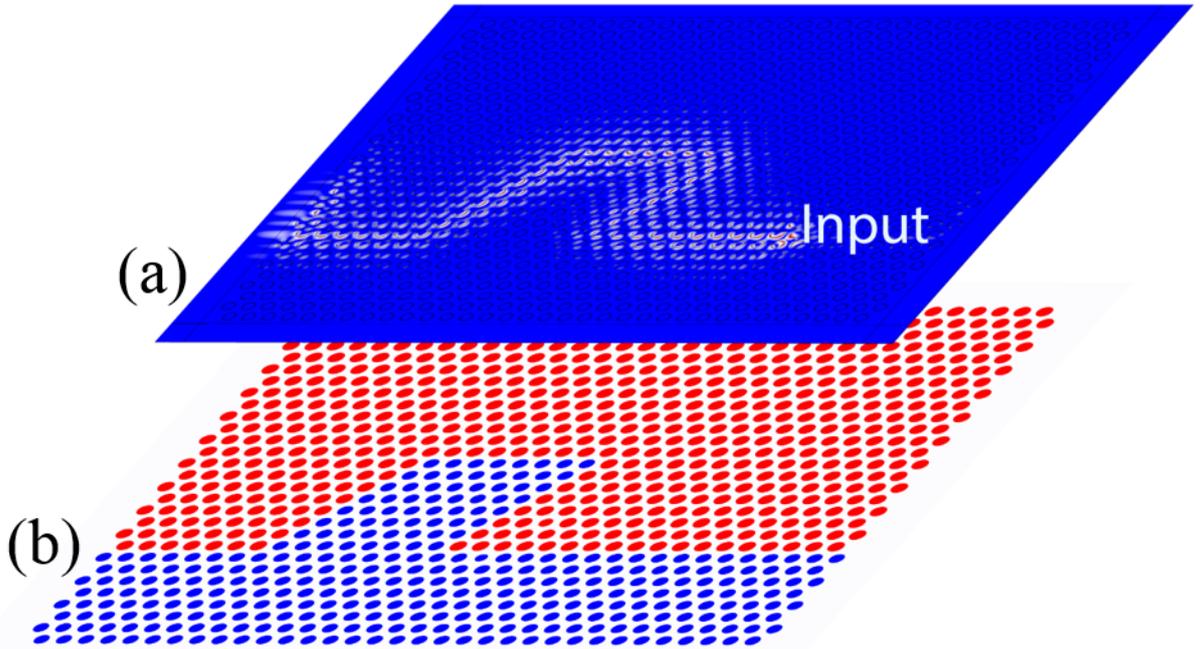

Figure 5. Unidirectional propagation of the edge states along an interface with sharp bends. The lower panel is a schematic of the sample in which the red dots represent the topological nontrivial phase while the blue dots indicate the topological trivial phase.

To conclude, we have designed a 2D acoustic crystal consisting of a triangular array of core-shell cylinders embedded in a water host. We have shown that a topological phase transition can be obtained by using the band inversion mechanism in such a simple system. A pseudo-time-reversal symmetry can be constructed by utilizing the $C_{6v}$ point group symmetry of the *p* and *d* eigenstates at the Γ point, and it follows that pseudo spin-up and spin-down states can be realized in this classical wave system in the same spirit as the quantum spin Hall effect in electronic systems. An effective Hamiltonian is developed for the current acoustic system around the Γ point, and it unveils the underlying mechanism that links the band inversion to a topological phase transition. Numerical simulations unambiguously demonstrate the unidirectional propagation of the pseudo-spin states and the robustness of these edge states against sharp bends. The simple design of our proposed structure suggests that experimental realization is feasible. Since the underlying principle is valid for different scales, we expect that it will have potential applications in manipulating and controlling acoustic waves over a very large frequency range, spanning from infrasound to ultrasound.

**Acknowledgements**

This work was supported by the National Natural Science Foundation of China (Grant Nos.11274120 and 11574087), the Fundamental Research Funds for the Central Universities (Grant No. 2014ZG0032), and King Abdullah University of Science and Technology (KAUST).